\documentclass{article}

\usepackage[final, nonatbib]{neurips_2023}

\usepackage[utf8]{inputenc} 
\usepackage[T1]{fontenc}    
\usepackage{hyperref}       
\usepackage{url}            
\usepackage{booktabs}       
\usepackage{amsfonts}       
\usepackage{nicefrac}       
\usepackage{microtype}      
\usepackage{xcolor}         
\usepackage{mathtools}
\usepackage{float}

\usepackage[style=ieee, uniquelist=false,
mincitenames=1,
maxcitenames=2,
minbibnames=1,
maxbibnames=2,
url=false,
doi=false,
eprint=false, 
backend=bibtex]{biblatex}

\title{Self-Supervised Music Source Separation Using Vector-Quantized Source Category Estimates}

\author{
  Marco Pasini$^*$ \\
  Queen Mary University \\
  London, UK \\
  \And
  Stefan Lattner$^*$ \\
  Sony Computer Science Laboratories\\
  Paris, France \\
  \And
  György Fazekas \\
  Queen Mary University \\
  London, UK \\
}

\begin{document}

\maketitle

\def\thefootnote{*}\footnotetext{Equal contribution}\def\thefootnote{\arabic{footnote}}

\begin{abstract}
  Music source separation is focused on extracting distinct sonic elements from composite tracks. Historically, many methods have been grounded in supervised learning, necessitating labeled data, which is occasionally constrained in its diversity. More recent methods have delved into N-shot techniques that utilize one or more audio samples to aid in the separation. However, a challenge with some of these methods is the necessity for an audio query during inference, making them less suited for genres with varied timbres and effects. This paper offers a proof-of-concept for a self-supervised music source separation system that eliminates the need for audio queries at inference time. In the training phase, while it adopts a query-based approach, we introduce a modification by substituting the continuous embedding of query audios with Vector Quantized (VQ) representations. Trained end-to-end with up to N classes as determined by the VQ's codebook size, the model seeks to effectively categorise instrument classes. During inference, the input is partitioned into N sources, with some potentially left unutilized based on the mix's instrument makeup. This methodology suggests an alternative avenue for considering source separation across diverse music genres. We provide examples and additional results online \footnote{\url{https://anonymous2732.github.io/vqss/}}.
\end{abstract}

\section{Introduction}

Source separation is a crucial domain in signal processing, aiming to extract individual components from mixed signals. This challenge has been widely addressed in the audio domain across various fields, including universal source separation, (multi-)speaker separation, and music source separation (MSS). MSS stands as a significant sub-domain, with the goal of isolating individual instruments, vocals, or other sonic elements from a composite track. Traditionally, MSS has been driven by supervised methods that necessitate large amounts of labeled stems. However, dedicated, openly available datasets are rare, and limited in the number of instrument classes. As a result, the default benchmark in MSS only involves four classes: vocals, drums, bass, and other. 

In order to lift this limitation, methods for unsupervised, semi-supervised, or self-supervised source separation have been investigated in recent years. A particular trend in semi- and self-supervised methods involves zero- one- or few-shot learning approaches. These models typically rely on one or multiple audio samples from a specific instrument class to guide the separation process. In practice, it is easier to find datasets of unlabelled or weakly labeled musical sounds, contained in public collections like AudioSet \cite{DBLP:conf/icassp/GemmekeEFJLMPR17}, Freesound \cite{DBLP:conf/mm/FontRS13}, or proprietary datasets. Consequently, using audio queries instead of instrument labels makes it possible to train self-supervised source separation models on such datasets (e.g., by using different segments of a stem as query and separation target).

However, the necessity of having the respective audio query available at inference time is a limiting factor in N-shot source separation. Particularly in contemporary popular or electronic music, the constituents can have peculiar timbres, like different synthesizers, guitar amplifiers, unusual samples, or sound effects. Therefore, ideally, a source separation system should be able to separate different instrument classes at inference time without requiring additional information.

This work introduces a proof-of-concept self-supervised music source separator that does not require audio queries at inference time. The method is a self-supervised query-based source separation system at train time, but we replace the continuous embedding space of the query audios with VQ embeddings. The model is trained end-to-end using a maximum of N classes, which is the codebook size of the VQ layer, to learn a helpful clustering of the respective instrument classes. During inference, the model divides the input mix into N sources. Depending on the presence of the different instrument classes in the mix, some outputs may remain empty.

\section{Method}
\label{method}

We propose a self-supervised MSS system that does not require audio queries during inference. The model is trained in a query-based fashion using random crops of stems as queries and targets. However, instead of using a continuous embedding space for the queries, we introduce a vector quantization layer to categorize the instrument classes into a discrete codebook.
The overall architecture consists of three main components: A \textit{Style Encoder} $E$ that embeds the audio query into a latent space, a \textit{Vector Quantizer} $Q$ that quantizes the continuous embedding into a discrete codebook, and a \textit{Generator} $G$ that generates the separated source conditioned on the quantized code.

Let $x$ be an unlabelled audio stem. We first randomly crop $x$ two times to obtain a target excerpt $x_{targ}$ and a reference excerpt $x_{ref}$. The target excerpt is linearly mixed with $k \sim \mathcal{U}(0,K)$ additional random stems to obtain the mixture $x_{mix}$.
The reference excerpt $x_{ref}$ is fed into the Style Encoder $E$ to produce a $d$-dimensional embedding $z_{ref} = E(x_{ref})$. This continuous embedding is then quantized by a Vector Quantization layer $Q$ to obtain a quantized embedding $\hat{z}_{ref} = Q(z_{ref})$ corresponding to the closest entry in a codebook of size $N$, representing the maximum number of different sources allowed by the system. The Generator $G$ takes as input the mixture $x_{mix}$ and the quantized embedding $\hat{z}_{ref}$, and is tasked with separating the target source $x_{targ}$ from the mixture:
\begin{equation*}
    \hat{x}_{targ} = G(x_{mix}, \hat{z}_{ref}).
\end{equation*}

We parameterize the Generator with a U-Net \cite{unet} model which takes magnitude spectrograms as input and outputs magnitude and phase spectrograms which are then combined into a waveform sample via the iSTFT operator. The model is inspired by the Autoencoder proposed in \cite{musika} and is thus trained in a similar fashion. Differently from \cite{musika}, the model is trained fully end-to-end using the following objectives: The \textit{Reconstruction loss} uses the same reconstruction objective \(\mathcal{L}_{rec}\) proposed in \cite{musika}, which consists of a L1 loss between log-spectrograms and a multi-scale spectral loss \cite{ddsp,rave} between the output \(\hat{x}_{targ}\) and target \(x_{targ}\). Details about the reconstruction objective can be found in the original paper. The \textit{Adversarial loss} utilizes the hinge adversarial loss \(\mathcal{L}_{adv}\) \cite{geometricgan} to implicitly model the phase spectrogram as:
\vspace{-0.7cm}

    \begin{align*}
    \mathcal{L}_{adv} = \phantom{+} &\mathbb{E}_{x_{targ}}[\min(0, -1 + D(\log(|\text{STFT}(x_{targ})|)))] \\
                      + &\mathbb{E}_{\hat{x}_{targ}}[\min(0, -1 - D(\log(|\text{STFT}(\hat{x}_{targ})|)))],
\end{align*}
where \(D\) is a Discriminator model. Lastly, the \textit{Commitment loss} is the VQ-specific objective \(\mathcal{L}_{vq}\) that ensures the encoder outputs are close to the quantized codes from the codebook: \(\mathcal{L}_{vq} = ||z_{ref} - \text{sg}(\hat{z}_{ref})||_2^2\), with \(\text{sg}\) denoting the stop-gradient operator. The codebook loss is realized through exponential moving average (EMA) updates, as suggested in \cite{vqvae,vqvae2}.

The overall training loss is a weighted sum of these objectives:
\begin{equation*}
    \mathcal{L} = \mathcal{L}_{adv} + \lambda_{rec}\mathcal{L}_{rec}  + \lambda_{vq}\mathcal{L}_{vq}.
\end{equation*}
At inference time, the model takes as input the mixture $x_{mix}$ and each of the final quantized embeddings and iteratively outputs $N$ estimated sources, without requiring any audio query. Depending on the actual instruments present in the mixture, some of the outputs may be silent.

This approach eliminates the need for audio queries during inference. The VQ-based categorization acts as an instrument classifier that is learned in a self-supervised manner through the source separation task. Furthermore, by training fully end-to-end, the model learns a discrete embedding space specifically tailored for the task of source separation.

\section{Implementation Details}
\label{impl}
The Generator, Style Encoder and Critic are fully convolutional models using standard residual blocks \cite{resnet} with group normalization layers \cite{groupnorm} with 8 groups. The Generator is based on a U-Net architecture where the conditioning information is introduced via FiLM layers \cite{film} at every block. The Style Encoder takes mel-spectrograms with 128 frequency bins as input and produces a single $512$-dimensional vector via a Global Average Pooling (GAP) layer. The VQ layer is initialized with $N=16$ embeddings via $10$ iterations of k-means clustering of the first input training batch. As proposed by \cite{improvedvqgan}, we use cosine similarity instead of Euclidean distance and we factorize the codes in a lower dimensional space with dimensionality of $8$ to improve codebook usage. The codebook loss is implemented via EMA with a factor of $0.99$. Regarding the STFT parameters used during training, we adopt the ones proposed by \cite{musika}. We choose $\lambda_{rec}=2.5$ and $\lambda_{vq}=100$. During training we use random crops of $1.5$ seconds with a sampling rate of $44.1$\,kHz to train the system. While we ensure that all $x_{ref}$ crops are not completely silent, we allow $x_{targ}$ crops to be silent, since it allows the model to produce a silent sample when the quantized embedding of a source that is not present in the input mix is used.
We train the model for 1 million iterations with a batch size of $32$, using Adam \cite{adam} with $\beta_1 = 0.5$ and $\beta_2 = 0.9$ as the optimizer.

\begin{figure}
    \centering
    \begin{minipage}{0.59\textwidth}
        \centering
        \includegraphics[width=1\linewidth]{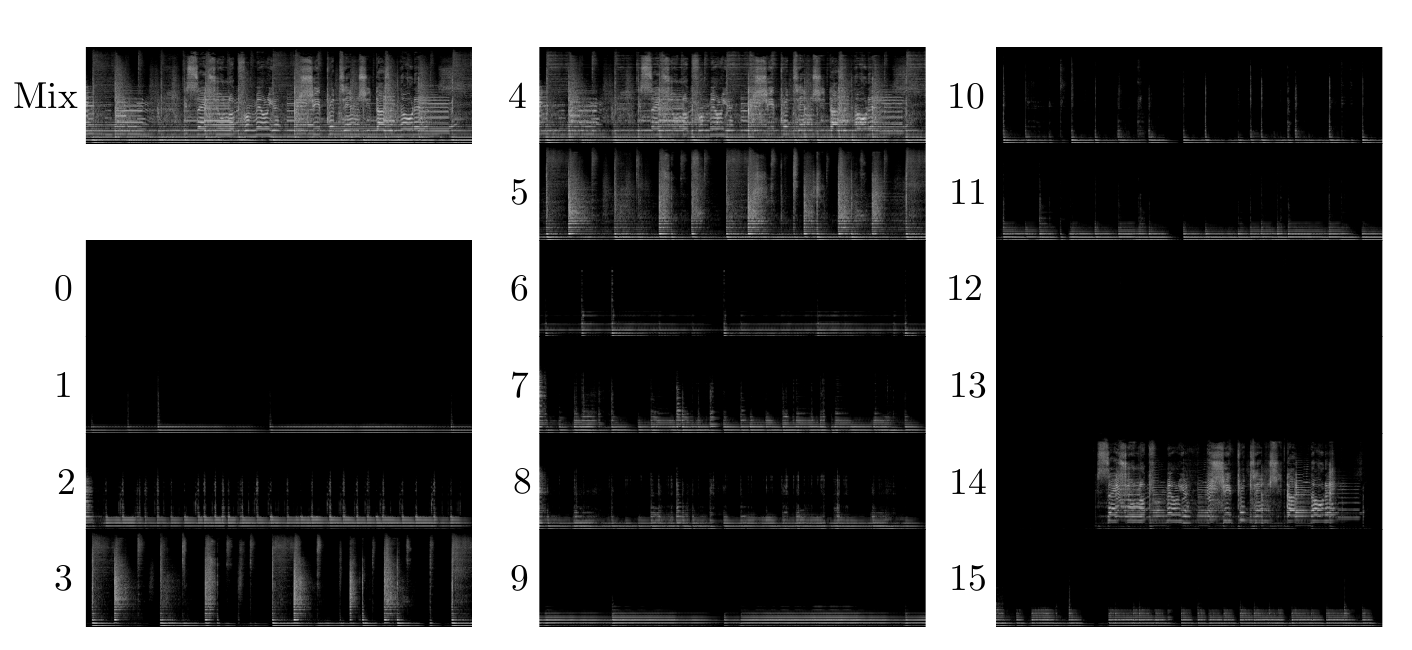}
        \vspace{-0.3cm}
        \caption{Mel-spectrograms of an example input mix and separated outputs. For each separation from 0 to 15 the Generator is conditioned on the corresponding quantized embedding. By cross-referencing this visualization with the histogram in Figure \ref{fig:hist}, it is possible to recognize specific instruments, such as \textit{Vocals} for cluster number 14.}
        \label{fig:specs}
    \end{minipage}%
    \hspace{0.02\textwidth}
    \begin{minipage}{0.39\textwidth}
        \centering
        \includegraphics[width=1\linewidth]{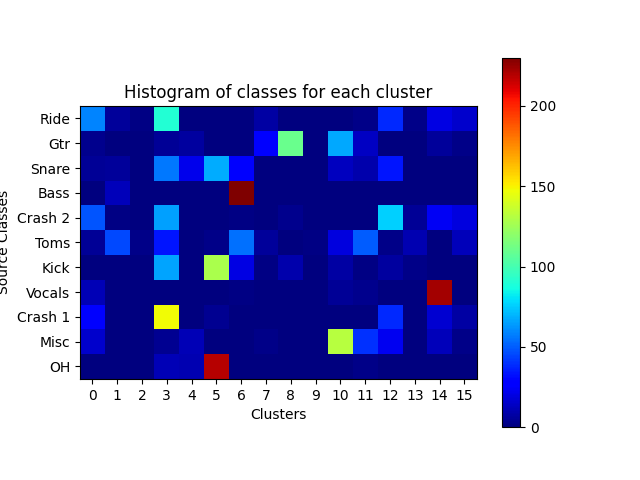}
        \vspace{-0.7cm}
        \caption{Distribution of the different classes of the test set for each quantized embedding found by the system. We notice a clear clustering of certain sources (\textit{OH, Bass, Vocals}), while for other classes (\textit{Toms, Snare, Ride}) the distribution is shared across different clusters.}
        \label{fig:hist}
    \end{minipage}
    \vspace{-0.3cm}
\end{figure}


\section{Experiments and Results}
\label{exps}

\begin{table}[h]
\footnotesize 
\centering
\scalebox{0.96}{
\begin{tabular}{|c|c|c|c|c|c|c|c|c|c|c|c|}
\hline
 & OH & Misc & Crash 1 & Vocals & Kick & Toms & Crash 2 & Bass & Snare & Gtr & Ride \\
\hline
L1 & 14.51 & 16.55 & 19.30 & 15.06 & 24.84 & 23.60 & 19.39 & 16.68 & 15.70 & 14.98 & 17.75 \\
\hline
L1 rand & 23.64 & 22.47 & 22.92 & 22.25 & 23.85 & 31.14 & 23.97 & 22.61 & 23.18 & 23.40 & 21.65 \\
\hline
\end{tabular}}
\caption{L1 distance between log-spectrograms of separated and true sources for each class in the test set. For comparison, we also calculate the L1 distance when the model is conditioned on a random quantized embedding instead of the target one. The difficulty of the model to correctly separate 'Kick' from other drums results in high L1 for that class.}
\label{table:eval}
\end{table}

We train the system on an internal collection of ~20,000 unlabelled stems. To evaluate the separation performance, we use an internal collection of 1,000 labelled stems with the following classes: \textit{OH, Misc, Crash 1, Crash 2, Vocals, Kick, Toms, Bass, Snare, Gtr, Ride}. 
We first visualize in Figure \ref{fig:hist} the distribution of the different classes for each final quantized embedding, or cluster, found by the system.
It is worth noting how the data distribution of training samples differs from the limited labelled collection used for testing. For example, training stems may not contain clearly separated percussive instruments, despite them being separated in different classes for the testing samples. This can explain the sub-par clustering performance on some specific classes used for testing.
Considering the specific capability of the system to perform separation of arbitrary learned sources, evaluation using standard metrics in the field of source separation \cite{sisdr} can be challenging, since different clusters can represent a single instrument class, and some uncommon instrument classes can be excluded from the clustering altogether.
In Table \ref{table:eval} we evaluate the separation performance of the model in terms of L1 distance between generated and target log-spectrograms for each class in the test set. More specifically, different crops serve as the target sample, which is mixed with 4 other sources to create the input mixture, and as the input to the Style Encoder to produce the quantized embedding used as conditioning. For comparison, we also calculate L1 distances when a random quantized embedding is used as conditioning.
We also visualize in Figure \ref{fig:specs} mel-spectrograms of an example test input mix and of the corresponding separations performed by the system, one for each of the codebook embeddings. By cross-referencing Figure \ref{fig:hist}, we can notice specific instruments being separated. However, we can also notice hallucination behavior that is characteristic of the proposed system: considering that we explicitly train a generative model to perform the task, the Generator may produce realistic separated outputs that are not actually present in the input mix.

\section{Related Work}
\label{sec:related_work}

Supervised neural network-based MSS has been tackled with multi-layer perceptrons \cite{DBLP:conf/eusipco/NugrahaLV16, DBLP:conf/icassp/UhlichGM15, DBLP:conf/waspaa/TakahashiM17}, CNNs and RNNs (and combinations thereof) \cite{DBLP:journals/corr/abs-2010-01733, DBLP:conf/icassp/UhlichPGEKTM17, DBLP:conf/iwaenc/TakahashiGM18, DBLP:journals/jossw/StoterULM19, DBLP:journals/taslp/LuoY23}, UNets \cite{DBLP:conf/ismir/StollerED18, DBLP:conf/icassp/Choi0CJ21} and with Transformers \cite{DBLP:journals/corr/abs-2309-02612}. In such supervised approaches, a specific model is typically trained for each source type.
Self-supervised paradigms involve permutation-invariant training for universal source separation \cite{DBLP:conf/icassp/YuKT017}, and an explicit instrument-clustering approach for MSS \cite{DBLP:conf/icassp/ChenWGR23}. 
Unsupervised approaches to source separation have been investigated using mixture-invariant training \cite{DBLP:conf/nips/WisdomTEWWH20} and unsupervised blind source separation with Variational Autoencoders \cite{DBLP:conf/eusipco/NeriBD21, DBLP:journals/taslp/LiKM23}. Audio query-based (i.e., N-shot) approaches exist in supervised fashions for one-shot \cite{DBLP:conf/ismir/LeeCL19} and few-shot \cite{DBLP:conf/icassp/WangSBB22} learning. Such query-based approaches are suitable for self-supervised training, where two segments of a stem are typically used as a separation target and a query, leading to zero-shot \cite{DBLP:conf/ismir/LinXKJ21} and one-shot \cite{DBLP:conf/icassp/GfellerRT21} architectures. Such methods typically use a single separation network with conditioning that is derived from embeddings of an audio query. Such a single separation model is also used in a meta-learning approach \cite{DBLP:conf/icassp/SamuelGN20} and for universal source separation with weakly labelled data \cite{DBLP:journals/corr/abs-2305-07447}, the latter uses conditioning from a separate audio tagging model.
The model proposed in this work is a one-hot conditioned UNet like in \cite{DBLP:conf/icassp/Choi0CJ21} that is trained in a self-supervised one-shot fashion like in \cite{DBLP:conf/icassp/GfellerRT21}. In addition, we perform an unsupervised clustering to inform the separation module like in \cite{DBLP:conf/icassp/ChenWGR23}. However, in contrast to \cite{DBLP:conf/icassp/ChenWGR23}, we perform this clustering in an end-to-end fashion using a VQ embedding space.

\section{Conclusion}
\label{conclusion}
This work introduced a proof-of-concept for a self-supervised MSS system that does not require audio queries during inference. The method adopts a fully end-to-end query-based approach during training, using random crops of stems as queries and targets. However, instead of a continuous embedding space, we introduced a vector quantization layer to categorize unlabelled sources into a discrete codebook. We demonstrate both the potential and current limitations of the proposed approach. On one hand, the model shows a clear ability to cluster certain common instrument classes fairly consistently. This indicates the feasibility of learning meaningful representations in a completely self-supervised manner using the source separation task as a proxy and without the explicit use of contrastive learning techniques.
On the other hand, we also highlight some flaws that need to be addressed in future work. Firstly, the model can exhibit a hallucination behavior, where it tends to generate realistic outputs even for sources that are not present in the input mixture. While this phenomenon is not uncommon in adversarial networks, it is undesirable for source separation where the objective is to filter out existing sources. Hallucinations could be alleviated by training on a dataset with a more balanced class distribution or by increasing the weight of the reconstruction loss. Secondly, the perceived audio quality of the separated outputs is currently lacking, with noticeable artifacts. This suggests that the model has not learned a robust generalized representation of the diverse instrument classes.

Overall, this work should be considered a first step towards fully self-supervised source separation models that do not rely on audio queries at test time. The clustering-based separation indicates a promising direction, but more research is required to improve the fidelity of the outputs and reduce hallucination behavior. With further development, end-to-end self-supervised source separation could provide flexibility to handle the wide range of sounds that occur in real-world music.

\printbibliography

@inproceedings{vqvae,
  author    = {A{\"{a}}ron van den Oord and
               Oriol Vinyals and
               Koray Kavukcuoglu},
  title     = {Neural Discrete Representation Learning},
  booktitle = {NeurIPS},
  year      = {2017}
}

@inproceedings{vqvae2,
  author    = {Ali Razavi and
               A{\"{a}}ron van den Oord and
               Oriol Vinyals},
  title     = {Generating Diverse High-Fidelity Images with {VQ-VAE-2}},
  booktitle = {NeurIPS},
  year      = {2019}
}

@inproceedings{adam,
  author    = {Diederik P. Kingma and
               Jimmy Ba},
  title     = {Adam: {A} Method for Stochastic Optimization},
  booktitle = {ICLR},
  year      = {2015}
}

@inproceedings{resnet,
  author    = {Kaiming He and
               Xiangyu Zhang and
               Shaoqing Ren and
               Jian Sun},
  title     = {Deep Residual Learning for Image Recognition},
  booktitle = {CVPR},
  year      = {2016}
}

@inproceedings{unet,
  author    = {Olaf Ronneberger and
               Philipp Fischer and
               Thomas Brox},
  title     = {U-Net: Convolutional Networks for Biomedical Image Segmentation},
  booktitle = {MICCAI},
  year      = {2015}
}

@inproceedings{DBLP:conf/icassp/YuKT017,
  author       = {Dong Yu and
                  Morten Kolb{\ae}k and
                  Zheng{-}Hua Tan and
                  Jesper Jensen},
  title        = {Permutation invariant training of deep models for speaker-independent multi-talker speech separation},
  booktitle    = {{ICASSP}},
  pages        = {241--245},
  publisher    = {{IEEE}},
  year         = {2017}
}

@inproceedings{DBLP:conf/nips/WisdomTEWWH20,
  author       = {Scott Wisdom and
                  Efthymios Tzinis and
                  Hakan Erdogan and
                  Ron J. Weiss and
                  Kevin W. Wilson and
                  John R. Hershey},
  title        = {Unsupervised Sound Separation Using Mixture Invariant Training},
  booktitle    = {NeurIPS},
  year         = {2020}
}

@inproceedings{DBLP:conf/eusipco/NeriBD21,
  author       = {Julian Neri and
                  Roland Badeau and
                  Philippe Depalle},
  title        = {Unsupervised Blind Source Separation with Variational Auto-Encoders},
  booktitle    = {{EUSIPCO}},
  pages        = {311--315},
  publisher    = {{IEEE}},
  year         = {2021}
}

@article{DBLP:journals/taslp/LiKM23,
  author       = {Li Li and
                  Hirokazu Kameoka and
                  Shoji Makino},
  title        = {FastMVAE2: On Improving and Accelerating the Fast Variational Autoencoder-Based Source Separation Algorithm for Determined Mixtures},
  journal      = {{IEEE} {ACM} Trans. Audio Speech Lang. Process.},
  volume       = {31},
  pages        = {96--110},
  year         = {2023}
}

@inproceedings{DBLP:conf/eusipco/NugrahaLV16,
  author       = {Aditya Arie Nugraha and
                  Antoine Liutkus and
                  Emmanuel Vincent},
  title        = {Multichannel music separation with deep neural networks},
  booktitle    = {{EUSIPCO}},
  pages        = {1748--1752},
  publisher    = {{IEEE}},
  year         = {2016}
}

@inproceedings{DBLP:conf/icassp/UhlichGM15,
  author       = {Stefan Uhlich and
                  Franck Giron and
                  Yuki Mitsufuji},
  title        = {Deep neural network based instrument extraction from music},
  booktitle    = {{ICASSP}},
  pages        = {2135--2139},
  publisher    = {{IEEE}},
  year         = {2015}
}

@inproceedings{DBLP:conf/waspaa/TakahashiM17,
  author       = {Naoya Takahashi and
                  Yuki Mitsufuji},
  title        = {Multi-Scale multi-band densenets for audio source separation},
  booktitle    = {{WASPAA}},
  pages        = {21--25},
  publisher    = {{IEEE}},
  year         = {2017}
}

@inproceedings{DBLP:conf/icassp/UhlichPGEKTM17,
  author       = {Stefan Uhlich and
                  Marcello Porcu and
                  Franck Giron and
                  Michael Enenkl and
                  Thomas Kemp and
                  Naoya Takahashi and
                  Yuki Mitsufuji},
  title        = {Improving music source separation based on deep neural networks through data augmentation and network blending},
  booktitle    = {{ICASSP}},
  pages        = {261--265},
  publisher    = {{IEEE}},
  year         = {2017}
}

@inproceedings{DBLP:conf/iwaenc/TakahashiGM18,
  author       = {Naoya Takahashi and
                  Nabarun Goswami and
                  Yuki Mitsufuji},
  title        = {{MMDenseLSTM}: An Efficient Combination of Convolutional and Recurrent
                  Neural Networks for Audio Source Separation},
  booktitle    = {{IWAENC}},
  pages        = {106--110},
  publisher    = {{IEEE}},
  year         = {2018}
}

@article{DBLP:journals/corr/abs-2309-02612,
  author       = {Wei Tsung Lu and
                  Ju{-}Chiang Wang and
                  Qiuqiang Kong and
                  Yun{-}Ning Hung},
  title        = {Music Source Separation with Band-Split RoPE Transformer},
  journal      = {CoRR},
  volume       = {abs/2309.02612},
  year         = {2023}
}

@inproceedings{DBLP:conf/icassp/ChenWGR23,
  author       = {Ke Chen and
                  Gordon Wichern and
                  Fran{\c{c}}ois G. Germain and
                  Jonathan Le Roux},
  title        = {Pac-HuBERT: Self-Supervised Music                 Source Separation
                  Via Primitive Auditory Clustering And Hidden-Unit Bert},
  booktitle    = {{ICASSP} Workshops},
  pages        = {1--5},
  publisher    = {{IEEE}},
  year         = {2023}
}

@article{DBLP:journals/corr/abs-2305-07447,
  author       = {Qiuqiang Kong and
                  Ke Chen and
                  Haohe Liu and
                  Xingjian Du and
                  Taylor Berg{-}Kirkpatrick and
                  Shlomo Dubnov and
                  Mark D. Plumbley},
  title        = {Universal Source Separation with Weakly Labelled Data},
  journal      = {CoRR},
  volume       = {abs/2305.07447},
  year         = {2023}
}

@inproceedings{DBLP:conf/ismir/LeeCL19,
  author       = {Jie Hwan Lee and
                  Hyeong{-}Seok Choi and
                  Kyogu Lee},
  title        = {Audio Query-based Music Source Separation},
  booktitle    = {{ISMIR}},
  pages        = {878--885},
  year         = {2019}
}

@inproceedings{DBLP:conf/ismir/LinXKJ21,
  author       = {Liwei Lin and
                  Gus Xia and
                  Qiuqiang Kong and
                  Junyan Jiang},
  title        = {A unified model for zero-shot music source separation, transcription
                  and synthesis},
  booktitle    = {{ISMIR}},
  pages        = {381--388},
  year         = {2021}
}

@inproceedings{DBLP:conf/icassp/GfellerRT21,
  author       = {Beat Gfeller and
                  Dominik Roblek and
                  Marco Tagliasacchi},
  title        = {One-Shot Conditional Audio Filtering of Arbitrary Sounds},
  booktitle    = {{ICASSP}},
  pages        = {501--505},
  publisher    = {{IEEE}},
  year         = {2021}
}

@inproceedings{DBLP:conf/icassp/WangSBB22,
  author       = {Yu Wang and
                  Daniel Stoller and
                  Rachel M. Bittner and
                  Juan Pablo Bello},
  title        = {Few-Shot Musical Source Separation},
  booktitle    = {{ICASSP}},
  pages        = {121--125},
  publisher    = {{IEEE}},
  year         = {2022}
}

@inproceedings{DBLP:conf/icassp/SamuelGN20,
  author       = {David Samuel and
                  Aditya Ganeshan and
                  Jason Naradowsky},
  title        = {Meta-Learning Extractors for Music Source Separation},
  booktitle    = {{ICASSP}},
  pages        = {816--820},
  publisher    = {{IEEE}},
  year         = {2020}
}

@article{DBLP:journals/jossw/StoterULM19,
  author       = {Fabian{-}Robert St{\"{o}}ter and
                  Stefan Uhlich and
                  Antoine Liutkus and
                  Yuki Mitsufuji},
  title        = {Open-Unmix - {A} Reference Implementation for Music Source Separation},
  journal      = {J. Open Source Softw.},
  volume       = {4},
  number       = {41},
  pages        = {1667},
  year         = {2019}
}

@article{DBLP:journals/corr/abs-2010-01733,
  author       = {Naoya Takahashi and
                  Yuki Mitsufuji},
  title        = {D3Net: Densely connected multidilated DenseNet for music source separation},
  journal      = {CoRR},
  volume       = {abs/2010.01733},
  year         = {2020}
}

@inproceedings{DBLP:conf/icassp/Choi0CJ21,
  author       = {Woo{-}Sung Choi and
                  Minseok Kim and
                  Jaehwa Chung and
                  Soonyoung Jung},
  title        = {Lasaft: Latent Source Attentive Frequency Transformation For Conditioned
                  Source Separation},
  booktitle    = {{ICASSP}},
  pages        = {171--175},
  publisher    = {{IEEE}},
  year         = {2021}
}

@article{DBLP:journals/taslp/LuoY23,
  author       = {Yi Luo and
                  Jianwei Yu},
  title        = {Music Source Separation With Band-Split {RNN}},
  journal      = {{IEEE} {ACM} Trans. Audio Speech Lang. Process.},
  volume       = {31},
  pages        = {1893--1901},
  year         = {2023}
}

@inproceedings{DBLP:conf/ismir/StollerED18,
  author       = {Daniel Stoller and
                  Sebastian Ewert and
                  Simon Dixon},
  title        = {Wave-U-Net: {A} Multi-Scale Neural Network for End-to-End Audio Source
                  Separation},
  booktitle    = {{ISMIR}},
  pages        = {334--340},
  year         = {2018}
}

@inproceedings{musika,
  author       = {Marco Pasini and
                  Jan Schl{\"{u}}ter},
  title        = {Musika! Fast Infinite Waveform Music Generation},
  booktitle    = {{ISMIR}},
  pages        = {543--550},
  year         = {2022}
}

@article{geometricgan,
  title={Geometric gan},
  author={Lim, Jae Hyun and Ye, Jong Chul},
  journal={arXiv preprint arXiv:1705.02894},
  year={2017}
}

@inproceedings{ddsp,
  author       = {Jesse H. Engel and
                  Lamtharn Hantrakul and
                  Chenjie Gu and
                  Adam Roberts},
  title        = {{DDSP:} Differentiable Digital Signal Processing},
  booktitle    = {{ICLR}},
  publisher    = {OpenReview.net},
  year         = {2020}
}

@article{rave,
  title={RAVE: A variational autoencoder for fast and high-quality neural audio synthesis},
  author={Caillon, Antoine and Esling, Philippe},
  journal={arXiv preprint arXiv:2111.05011},
  year={2021}
}

@inproceedings{groupnorm,
  author       = {Yuxin Wu and
                  Kaiming He},
  title        = {Group Normalization},
  booktitle    = {{ECCV} {(13)}},
  series       = {Lecture Notes in Computer Science},
  volume       = {11217},
  pages        = {3--19},
  publisher    = {Springer},
  year         = {2018}
}

@inproceedings{film,
  author       = {Ethan Perez and
                  Florian Strub and
                  Harm de Vries and
                  Vincent Dumoulin and
                  Aaron C. Courville},
  title        = {FiLM: Visual Reasoning with a General Conditioning Layer},
  booktitle    = {{AAAI}},
  pages        = {3942--3951},
  publisher    = {{AAAI} Press},
  year         = {2018}
}

@inproceedings{improvedvqgan,
  author       = {Jiahui Yu and
                  Xin Li and
                  Jing Yu Koh and
                  Han Zhang and
                  Ruoming Pang and
                  James Qin and
                  Alexander Ku and
                  Yuanzhong Xu and
                  Jason Baldridge and
                  Yonghui Wu},
  title        = {Vector-quantized Image Modeling with Improved {VQGAN}},
  booktitle    = {{ICLR}},
  publisher    = {OpenReview.net},
  year         = {2022}
}

@inproceedings{sisdr,
  author       = {Jonathan Le Roux and
                  Scott Wisdom and
                  Hakan Erdogan and
                  John R. Hershey},
  title        = {{SDR} - Half-baked or Well Done?},
  booktitle    = {{ICASSP}},
  pages        = {626--630},
  publisher    = {{IEEE}},
  year         = {2019}
}

@inproceedings{DBLP:conf/icassp/GemmekeEFJLMPR17,
  author       = {Jort F. Gemmeke and
                  Daniel P. W. Ellis and
                  Dylan Freedman and
                  Aren Jansen and
                  Wade Lawrence and
                  R. Channing Moore and
                  Manoj Plakal and
                  Marvin Ritter},
  title        = {Audio Set: An ontology and human-labeled dataset for audio events},
  booktitle    = {{ICASSP}},
  pages        = {776--780},
  publisher    = {{IEEE}},
  year         = {2017}
}

@inproceedings{DBLP:conf/mm/FontRS13,
  author       = {Frederic Font and
                  Gerard Roma and
                  Xavier Serra},
  title        = {Freesound technical demo},
  booktitle    = {{ACM} Multimedia},
  pages        = {411--412},
  publisher    = {{ACM}},
  year         = {2013}
}

\end{document}